\documentclass[a4paper,twoside,twocolumn]{article}
\pdfoutput=1

\newif\iftr 
\trtrue     

\usepackage{ifpdf}  
\ifpdf
    \usepackage[pdftex]{graphicx}
    \usepackage[colorlinks,linkcolor=blue,citecolor=blue,urlcolor=blue]{hyperref}
    \DeclareGraphicsExtensions{.pdf}
\else
    \usepackage{graphicx}
    \usepackage[hypertex]{hyperref}  
\fi

\usepackage{todonotes} 
\newcommand{\bob}[1]{\todo[color=olive!40,inline]{Bob: #1}}

\usepackage{
url       %
,fancyhdr 
,lastpage 
,enumerate 
,amsmath  
,amsfonts 
,amssymb  
, enumitem 
}
\usepackage[hang]{footmisc} 
\usepackage{balance}

\usepackage[noindent, arraystretch, fullpage]{setlengths} 
\usepackage{
own         
}

\usepackage{amsthm}

\newtheorem*{test*}{Test}
\usepackage{xfrac}
\newcommand{\incr}{\mathrel{{+}{=}}}

\graphicspath{{images/}}

\newcommand*{\metaauthori}{Bob Briscoe}
\newcommand*{\metashorttitle}{DCTCP Clock Machinery Lag}
\newcommand*{\metatitle}{Removing the Clock Machinery Lag from DCTCP/Prague}
\newcommand*{\metano}{TR-BB-2020-002}
\newcommand*{\metakeywords}{Data Communications, Networks, Internet,
Data Centre, Performance, Latency, Delay, Responsiveness, Congestion, Control, Feedback, Smoothing, Filtering, Algorithm}

\newcommand*{\metamaili}{\href{mailto:research@bobbriscoe.net}{research@bobbriscoe.net}}
\newcommand*{\metaaddress}{}

\newcommand*{\metaversion}{04}
\newcommand*{\metadate}{7 Sep 2022}

\hypersetup{                       
     pdfauthor = {\metaauthori
     },
     pdftitle = {\metashorttitle},
     pdfsubject = {},
     pdfkeywords = {\metakeywords}
}%

\title{\metatitle}%
\author{\metaauthori%
\thanks{\metamaili, %
\metaaddress}%
}
\date{\metadate}%

\fancypagestyle{first}{%
\fancyhead[LO,RE]{}%
\fancyhead[LE,RO]{}%
\fancyhead[C]{}%
}%

\begin{document}
\bibliographystyle{alpha}%


\maketitle%
\thispagestyle{first}

\begin{abstract}
{\small\noindent%
This report explains how DCTCP takes 2--3 rounds before it even starts to
respond to congestion. This is due to the clocking machinery in its moving
average of congestion feedback. Instead, per-ACK mechanisms are proposed, which
cut out all the extra lag, leaving just the inherent single round of feedback
delay. Even though clocking per ACK updates the average much more frequently, it
is arranged to inherently smooth out variations over the same number of round
trips, independent of the number of ACKs per round. 

Evaluation of the v02 algorithm found design errors. This version (v04) is
published prior to evaluation, in order to elicit early feedback on the design.
}      
\end{abstract}
\section{Problem}\label{prresp_Problem}


Classic Active Queue Management (AQM), as in RED, CoDel or PIE, smooths out
rapid variations in the queue before the AQM signals any congestion. In DCTCP~\cite{Alizadeh10:DCTCP},
Alizadeh \emph{et al} introduced the idea of an immediate AQM, that signals
congestion without any filtering, then the sender smooths these signals using an
EWMA, before using the smoothed average to control the load it applies to the
network, via its congestion window.

In the classical approach, smoothing in the network introduces a worst-case round trip time
(RTT) of smoothing delay (about 100\,ms for AQMs used in the 
Internet), because it is hard for a network element to determine the actual RTTs
of all the flows traversing it. This is the `good queue' described by Nichols \& Jacobson~\cite{Nichols18:codel_RFC}.

However, when reducing delay, there is no such thing as a good queue. In the
DCTCP approach, a sender can tailor its smoothing delay to its own RTT, which it
measures anyway for other reasons, Then, it only introduces as much smoothing
delay as is necessary to smooth itself.

However, this report shows that common implementations of DCTCP (e.g.\ in
Windows, Linux, or FreeBSD~\cite{Bensley17:DCTCP}) add one to two rounds of lag
before regulating their load in response to congestion. This lag is on top of
the inherent round trip of delay in the feedback loop. It is in the
clocking machinery that ensures the feedback is smoothed over a set number of 
round trips. 

Derivatives of DCTCP inherit the same machinery delay; for instance the 
Prague congestion control~\cite{Briscoe21b:PragueCC-ID}, which is intended for
use over the wide area where unnecessary rounds of lag
will be much more noticeable in absolute terms. Unless otherwise stated, the
term DCTCP in this paper should be considered to include derivatives like
Prague.
\begin{figure}[h]
	\centering
		\vspace{-6pt}
	\includegraphics[width=0.9\linewidth]{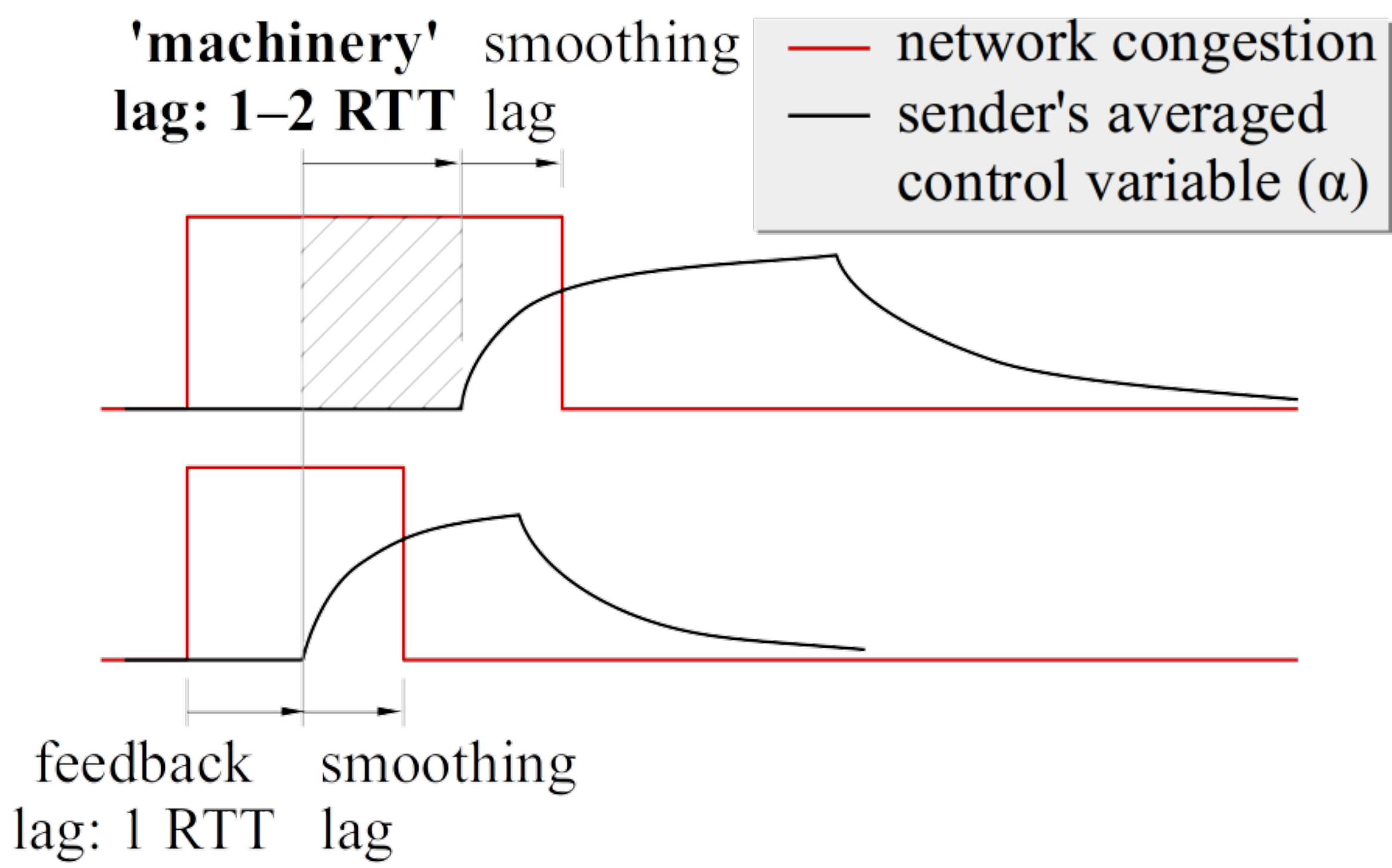}
	\vspace{-6pt}
	\caption{Schematic illustrating averaging of a congestion impulse at the sender
	before and after removing the `machinery' lag addressed in this
	report}\label{fig:lag-schematic}
	\vspace{-12pt}
\end{figure}
\begin{figure*}
	\includegraphics[width=\linewidth]{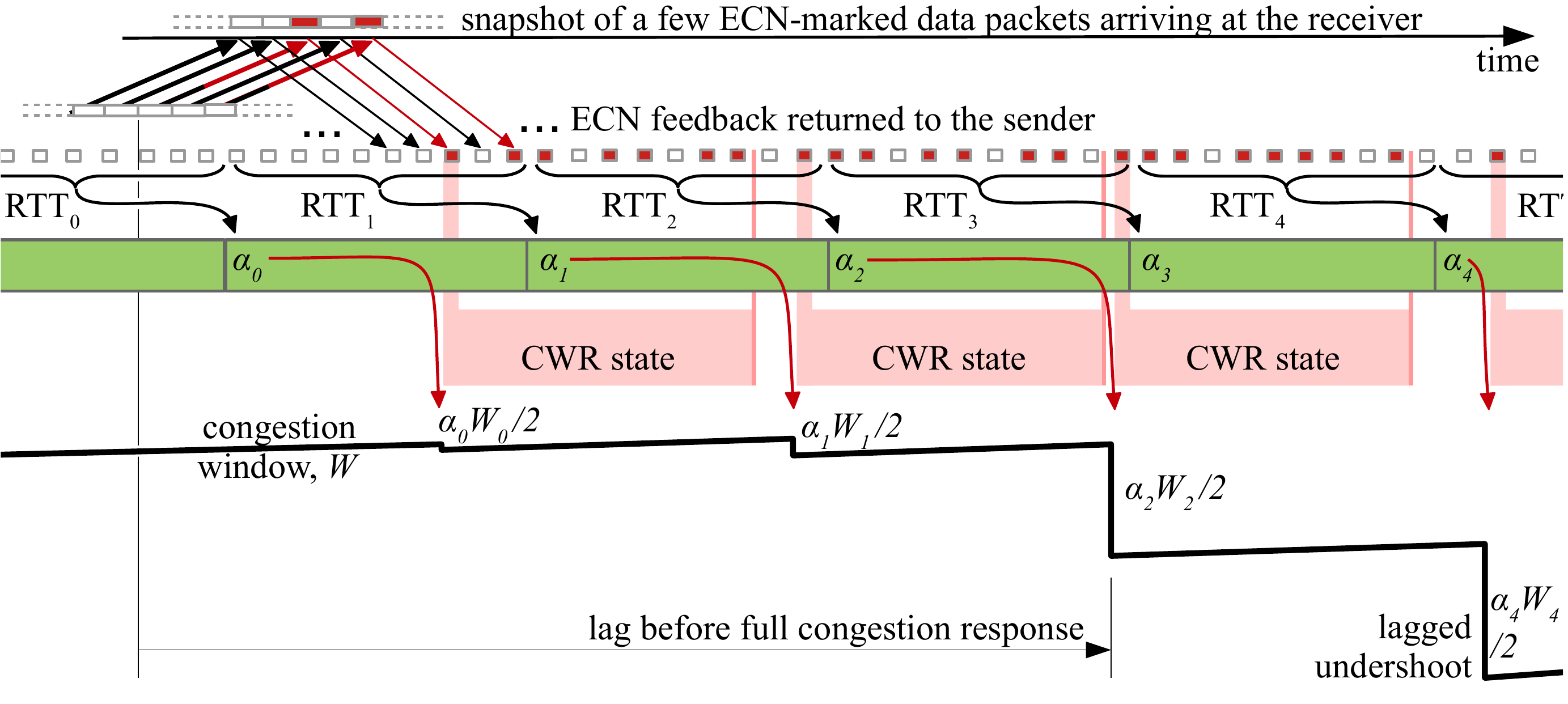}
	\caption{The problem: DCTCP's two stages for processing congestion feedback: 1)
		gathering feedback in a fixed sequence of rounds (RTT\(_i\)) to calculate the
		EWMA (\(\alpha_i\)); 2) applying this EWMA on the first feedback mark, when it
		has had no time to gather enough feedback, which leads to a typically
		inadequate congestion response before entering congestion window reduced (CWR)
		state, which suppresses any further response for a round. See text for full
		commentary.}
	\label{fig:dctcp-feedback-lag}
\end{figure*}

A moving average intentionally adds smoothing delay to dampen responsiveness 
in case a change does not sustain over the whole averaging period.
However, because smoothing spreads the response out, it only adds lag to the end, 
not to the start. In contrast, machinery delay represent pure lag before the
damping can even start (as illustrated in \autoref{fig:lag-schematic}).

This means that established DCTCP flows take 2--3 rounds (rather than one)
before they even start to respond to a reduction in available capacity or yield
to a new flow. This is likely to lead to more overshoot and undershoot of the 
control system, leading to more queue delay variation and less utilization than 
necessary. In turn, this means that a new flow must either build up a large
queue before any established flow yields or, to avoid excessive queuing, it 
must enter the system much more
tentatively than necessary. Therefore, the algorithm in this report is at least part of a
solution to the 'Prague Requirement' for 'faster convergence at flow
start'~\cite[Appx A.2.3]{Briscoe15f:ecn-l4s-id_ID}.

The extra rounds are due to DCTCP's two-stage process for responding to
congestion (see \autoref{fig:dctcp-feedback-lag}):
\begin{enumerate}
	\item The first stage (shown in green) introduces
	one round of delay (RTT\(_i\)) while it accumulates the marking fraction before
	it can calculate the EWMA (\(\alpha_i\)). It passes this to the second stage
	that reduces the congestion window (\(W\)) by \(\alpha_i W_i/2\). 
	\item The second extra
	round arises because the second stage is triggered by congestion feedback (a red
	ACKnowledgement in \autoref{fig:dctcp-feedback-lag}) that occurs independently
	of the regularly clocked first stage.
	This is exacerbated by entering congestion window reduced
	(CWR) state (shown in pink) at the first sign of congestion, which in turn suppresses any further
	response for a round --- just when more congestion feedback is likely. 
\end{enumerate}

So, after the one round of feedback delay, it takes up to two further rounds 
before a full round of the congestion that triggered the
start of the second stage has fed through into the EWMA that the first stage
passes to the second. Also, the
lagged congestion response will tend to overrun into the subsequent round,
causing undershoot.

DCTCP's current clocking machinery is built to ensure that its EWMA smooths over
a defined number of round trips, no matter the ack rate. To ensure stability%
\footnote{Actually DCTCP uses more smoothing than the minimum
	necessary for stability. This is so that long-running DCTCP flows will leave a
	degree of capacity headroom at the bottleneck for the prevailing level of short
	flows, which would otherwise cause spikes of queue variation.}%
, the damping of the control system is meant to
depend solely on the RTT, which is the only inherent lag in the system.
Otherwise, if the EWMA were updated on
every ACK, the more ACKs there were per round, the shorter the duration over
which the EWMA would smooth. 

So, the problem boils down to how to update an EWMA on every ACK, but
arrange for its value to evolve as if it is only updated once per round trip, even though
the number of ACKs per round varies. 

%
%

\section{Per-ACK Machinery Design}\label{prresp_Per-Packet_Machinery}\label{prresp_intuition}
\subsection{Definitions of variables}
\begin{description}[nosep]
	\item [\texttt{g}]: \((0 < \mathtt{g} < 1)\) the gain or relative weight given
    to new data over old;
	\item [\texttt{G}] = \texttt{1/g} (\(>1\)). Characteristic smoothing delay in
    update cycles. By default in DCTCP \texttt{G = 16};
	\item [\texttt{av\_up}]: EWMA of the marks per round upscaled by \texttt{G};
	Alternatively, it might help to think of this as the EWMA of the marking
	probability (alpha in DCTCP) upscaled by \texttt{G * acks\_}. 
	\item [\texttt{cwnd}]: no.\ of packets in the congestion window;
	\item [\texttt{flight}]: no.\ of packets in flight now;
	\item [\texttt{acks}]: no.\ of ACKs per round, used for dividing per-round
	changes into per-ACK increments;
	\item [\texttt{acks\_}]: no.\ of ACKs per round used for upscaling;%
	\footnote{Initially, \texttt{acks\_} \& \texttt{acks} are taken as the same,
		but see \S\,\ref{prresp_Variable_Upscaling}.}
	\item [\texttt{ce\_fb}]: no.\ of congestion-marked packets fed back on a particular ACK.
\end{description}
For simplicity, \texttt{cwnd}, \texttt{flight} and
\texttt{ce\_fb} are set in units of packets. However, units of bytes are preferable,
to guard against ACK-splitting attacks~\cite{Savage99:ECN_nonce} and to
correctly respond to congestion marking on different sized packets.

\subsection{Per-ACK EWMA}\label{prresp_Per-Packet_EWMA}

An EWMA of any variable \(x\) can be computed by repeated execution of
\vspace{-6pt}
\[\bar{x} \gets gx + (1-g)\bar{x}.\]
In integer arithmetic, the EWMA would usually be upscaled by \(1/g\),%
\footnote{This technique was used by Van Jacobson to implement the gain of TCP's
	smoothed RTT in integer arithmetic. It was also used to implement the upscaled
	EWMA of the congestion fraction (alpha) in TCP Prague (on top of the upscaling
	by 10 bits already applied in DCTCP). In these cases the upscaling was fixed.
	But the same principle applies for variable upscaling.}
otherwise multiplying it by the fraction \(g\) would lose precision:
\vspace{-6pt}
\[\overline{x_\mathrm{up}} \gets x + (1-g)\overline{x_\mathrm{up}}.\]
Rewriting this in terms of \(G\), and rearranging:
\vspace{-6pt}
\[\overline{x_\mathrm{up}} \incr x - \overline{x_\mathrm{up}}/G.\]
Whenever the upscaled EWMA is used, it has to be downscaled again, that is,
divided by \(G\).

To update the EWMA per-ACK rather than per-RTT, it is proposed 
to reduce the gain to \texttt{g / acks\_}, where
\texttt{acks\_} is the number of ACKs per round.%
\footnote{The approximation error is quantified in \autoref{prresp_approx}.} 
Usually the gain of an EWMA is constant, but \texttt{acks\_} varies, 
so this makes the gain variable.%
\footnote{Care is needed, because earlier values could have been fed into the
	EWMA when it was upscaled by a very different amount --- see
	\S\,\ref{prresp_Variable_Upscaling}.}
As just described, this would be implemented by upscaling the EWMA by 
\texttt{acks\_ * G}.

Per-ACK clocking produces almost the same result as DCTCP, but without the extra
complexity and lag of the machinery needed to explicitly clock the EWMA once
per-round.
Specifically, no matter how many ACKs there are per RTT, upscaling the EWMA by
\texttt{acks\_ * G} implicitly introduces almost the same characteristic
smoothing delay (\texttt{RTT * G}) as explicitly clocking once per round. 

The following table shows the extra upscaling that compensates for the faster
update frequency of the per-ACK EWMA, so that the smoothing time remains
unchanged at \(G\) round trips (right-hand column).

\begin{table}[h]
	\centering
	\begin{tabular}{ccrc}
	Updates       &                 &                 & Smoothing\\
	per RTT       &        Max incr &       Upscaling & time [RTT]\\
	\hline
	1             &               1 &               G & G\\
	\texttt{acks} &               1 & \texttt{acks\_}*G & G\\
	\texttt{acks} & \texttt{ce\_fb} & \texttt{acks\_}*G & G\\
	\end{tabular}
	\caption{Adjusting Upscaling to keep Smoothing Time Unchanged}\label{tab:invariant}
\end{table}

The first row represents DCTCP (or Prague). 
The value '1' in the `Max incr.' column indicates the maximum value fed into the
EWMA per update event. In DCTCP this is limited to 1, because it feeds in the
fraction of marked packets per round.

The second row represents an example per-ACK EWMA. Because there are more update
events per round, the gain has to be reduced to compensate, which is achieved
here by upscaling the EWMA by the number of ACKs per round. In this example, the
max incr. is still 1, which implies that, on each ACK, the fraction of marked
packets relative to all the packets covered by the ACK is fed into the EWMA.%
\footnote{This is prone to bias, which is why we use the third row (see 
	\S\,\ref{prresp_biased_acks}).} 
To extract the average fraction of marks from this EWMA, it would have to be
downscaled again, by dividing by \texttt{acks\_ * G}.

In the third row, the EWMA is incremented by \texttt{ce\_fb}, the number (not the
proportion) of marked packets fed back by each ACK. To extract the average
fraction of marks from this EWMA, it would have to be divided by \texttt{acks\_
	* pkt/ack * G} = \texttt{flight*G}. Alternatively, just dividing this EWMA by
\texttt{G} would give the EWMA of the number (not the proportion) of marks per
round.%
\footnote{Note that increasing the max value fed into the EWMA merely increases
	the numerical value of the EWMA, but it does not need to be compensated by more
	upscaling (for example, it would not be appropriate to replace \texttt{acks\_}
	by \texttt{flight} in the upscaling).}

This comparison with DCTCP will now be written in pseudocode. 
In DCTCP the EWMA of the proportion of marks,
\texttt{alpha}, is maintained per round trip as follows (in floating point
arithmetic):
\vspace{-12pt}
\begin{verbatim}
    alpha += (F - alpha)/G,
\end{verbatim}
\vspace{-12pt}
where F is the fraction of marked bytes accumulated over the last round trip.

In Prague, for integer arithmetic, this EWMA is upscaled to \texttt{alpha\_up} 
by not including \texttt{F} in
the division, as follows:
\vspace{-12pt}
\begin{verbatim}
    alpha_up += F - alpha_up / G.
\end{verbatim}

This per-RTT EWMA can be approximated (see \autoref{prresp_approx}) by repeatedly updating
the EWMA on the feedback of every ACK, but scaling down each update by the
number of ACKs in that round, \texttt{acks\_}. That is:
\vspace{-12pt}
\begin{verbatim}
    av_up += ce_fb - av_up/(acks_*G).
\end{verbatim}
The result, \texttt{av\_up}, can be thought of as either i) the EWMA of the
proportion of marks upscaled by \texttt{flight*G}; or ii) the EWMA of the number
of marks per round,
upscaled by just \texttt{G}.

\subsection{Per-ACK Congestion Response}\label{prresp_Per-ACK_Response}

On first onset of congestion, DCTCP immediately responds with a tiny reduction
(assuming absence of congestion for a while previously). But then, perversely,
it suppresses any further response for a round trip.
This suppression of any further response for a round mimics classical congestion
controls. However, the response of a classical
CC is large and fixed, so it makes sense to then hold back for a round because
the initial response will usually have been sufficient. In contrast, DCTCP's
initial response is extent-based and typically small. So it makes no sense to
then suppress any further response for a round, just when most congestion
feedback is likely to be appearing. It would only make sense for DCTCP to mimic
the timing of a Classic response if it also mimicked its size.

The approach proposed in this section is not necessarily the final word on how
to use the per-ACK EWMA for scalable congestion control (see
\S\,\ref{prresp_future} for further ideas). Nonetheless, as a first step, we
build incrementally on DCTCP, using its teaching selectively, but not straying
too far from its intent. 

First, as with DCTCP, once feedback of a CE mark triggers the start of a
congestion response, congestion window reduction (CWR) state is entered for 1
RTT, during which no further response will start. Then once CWR state ends,
another congestion response will only start if new CE feedback arrives at the
sender.

However, instead of one reduction at the start of CWR state, the reduction is
spread over the duration of CWR state, to exploit each update of the EWMA
marks (\texttt{av\_up}) now that it is continually updated per-ACK. 
So, at one extreme, if the first CE mark is immediately
followed by many others, the EWMA will rapidly increase early in the round of
CWR, and \texttt{cwnd} will be rapidly decreased accordingly. While, at the other
extreme, if the first CE mark is the only one in the round, \texttt{cwnd}
will still have reduced by \texttt{av\_up/(2*G)} by the end of the round, but
\texttt{av\_up} will hardly have increased above the value it took when the CWR round
started. 

This combination of the timing and size of the reduction 
keeps the `\(\sfrac{1}{p}\)' congestion response of DCTCP. That is, the timing
of the start of each reduction is the same as DCTCP, and the size of the spread
out reduction ends up roughly the same size as if DCTCP had updated alpha and
used it about a round trip after its (usually inadequate) first response.
Except, in the proposed approach, the response starts immediately on feedback 
of the first congestion mark,rather than being lagged by 1--2 rounds.

As before, this comparison with DCTCP will now be described in pseudocode. 
In response to congestion, DCTCP (and Prague) multiplicatively decrease the
window no more than once per round as follows:
\vspace{-12pt}
\begin{verbatim}
    cwnd -= alpha/2 * cwnd;
\end{verbatim}
\vspace{-12pt}
where alpha is the EWMA of the marking proportion.

As explained earlier, the per-ACK EWMA, \texttt{av\_up}, represents the EWMA of
the number (not proportion) of marks per round, upscaled by \texttt{G}.
So, a multiplicative decrease per-round would be implemented as follows:
\vspace{-12pt}
\begin{verbatim}
    cwnd -= av_up/(2*G);
\end{verbatim}
\vspace{-12pt}
However, instead of applying this reduction once, it is divided out over every
ACK during CWR state, whatever the feedback on each ACK (much as many
implementations divide out
their additive increase over a round), as follows:
\vspace{-12pt}
\begin{verbatim}
    cwnd -= av_up/(2*G*acks_);
\end{verbatim}
But \texttt{av\_up} is updated each time an ACK arrives, so the spread out
reduction picks up new values of the EWMA as it evolves.

Notice that there is no multiplication by \texttt{cwnd}, because
\texttt{av\_up/G} is the averaged \emph{number} of marks, which is broadly the
same as the averaged \emph{proportion} of marks multiplied by
\texttt{cwnd}.\footnote{But with a subtle difference; the marks that drive the
	window reduction represent a proportion of the window
actually \emph{used} and acknowledged, not the maximum that the flow was
entitled to use (\texttt{cwnd}). Thus, if an
application-limited flow has only used a quarter of the available window, the
proposed reduction will be only a quarter of
that which would be applied by DCTCP (see
\S\,\ref{prresp_Advantage_App-Limited}, which argues that this could be a good
thing. Nonetheless, if it turns out not to be, an adjustment for \texttt{cwnd}
can be made).}

\section{Implementation}\label{prresp_implementation}

\subsection{Maintaining the EWMA}

On every ACK event, the \texttt{ce\_fb} term can be implemented by adding the
number of marked packets fed back to \texttt{av\_up}. 
The number of packets acknowledged in the current round is \texttt{acks}. So
repeatedly subtracting \texttt{av\_up/(acks\_*G)} on the arrival of every ACK
would reduce \texttt{av\_up} by \texttt{av\_up*acks/(acks\_*G)} in a round.
This approximates to \texttt{av\_up/G} per round (see \autoref{prresp_approx}
and \S\,\ref{prresp_Variable_Upscaling}).

\begin{verbatim}
On_each_ACK'd_packet {
  acks_ = update_acks();
  // Update EWMA
  av_carry = div(
    av_carry.rem+av_up, acks_*G);
  av_up += ce_fb - av_carry.quot;
}
\end{verbatim}

First the value of \texttt{acks\_} is updated before it is used. Two fairly
straightforward alternative implementations of \texttt{update\_acks()} are given
in \S\,\ref{prresp_acks_}. 
The pseudocode uses an integer division function \texttt{div()} with the same
interface as \texttt{div()} in C's standard library.\footnote{\texttt{div()} can
be thought of as a wrapper round \texttt{do\_div()}, which is intended for
kernel use, but less readable for pseudocode.} Specifically, it takes numerator
and denominator as parameters and returns both the
quotient and the remainder in the following structure:
\begin{verbatim}
typedef struct {
  int quot;
  int rem;
} div_t;
\end{verbatim}
The variable \texttt{av\_carry} would be declared of type \texttt{div\_t}. It is
used to carry forward the remainder to the invocation on the next ACK. The
quotient will typically be either 0 or 1, which is then used to decrement the
EWMA.\footnote{Given the result will nearly always be 0, sometimes 1 and hardly
	ever more, it would be possible to implement this division as a couple of
	conditions to test for 0 and 1, then do the division otherwise.}

In practice, values would need to be checked for underflow and/or overflow (for
instance, \texttt{av\_up} has to be prevented from going negative in the last
line), but such details are omitted from the pseudocode in this paper.

\begin{figure}[h]
	\includegraphics[width=\linewidth]{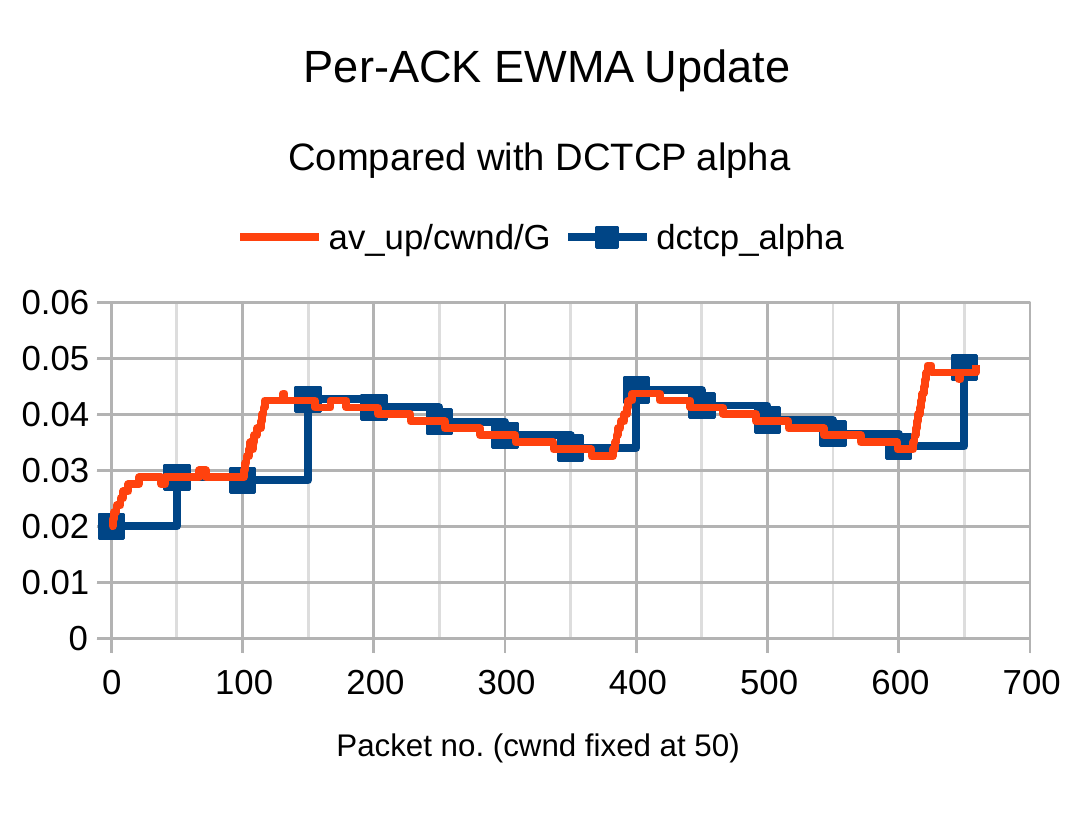}
	\caption{Initial verification of the first stage of the per-ACK EWMA 
	algorithm (with constant \texttt{cwnd}).}\label{fig:per-ack-ewma-verify}
\end{figure}

\autoref{fig:per-ack-ewma-verify} compares toy simulations of the above EWMA and
the DCTCP EWMA (both without changing cwnd). It can be seen that, whenever marks
arrive, the algorithm always moves immediately, whereas DCTCP's EWMA does
nothing until the next round trip cycle.

To initialize the EWMA, on the first CE marked feedback, \texttt{av\_up = cwnd * G}
would mimic the initialization of \texttt{alpha} in DCTCP and in the current
implementation of Prague. It might be worth
experimenting with possible improvements; for instance using \texttt{av\_up = flight * G}
instead.

\subsection{Responding to Congestion}\label{prresp_congestion_response}

As outlined in \S\,\ref{prresp_Per-ACK_Response}, a congestion reduction is divided%
\footnote{Note that the reason for the division by \texttt{acks} here is to
	divide the reduction over the ACKs in the round, whereas the previous division
	by \texttt{acks\_} (note the trailing underscore) was to upscale the EWMA.}
over a whole round of ACKs during CWR state, which reduces \texttt{cwnd} as the
EWMA updates, such that, by
the end of the round it will still have reduced roughly as much as it would have done if
the whole reduction had been applied at the end:
\vspace{-12pt}
\begin{verbatim}
    cwnd -= av_up / (acks*2*G);
\end{verbatim}
\vspace{-12pt}

To spread the reduction over the round, the proposed algorithm below does not
divide the round into an arbitrary number of points where \texttt{cwnd} is
altered by varying amounts. Instead, it calculates how much to carry on every ACK
and decrements \texttt{cwnd} every time at least one packet of movement is possible. 

\begin{verbatim}
On_each_ACK'd_packet {
  acks_ = update_acks();
  // Update EWMA
  av_carry = div(
    av_carry.rem+av_up, acks_*G);
  av_up += ce_fb - av_carry.quot;

  if (!cwr && ce_fb) {
    // Record start of CWR state
    next_seq = snd_next;
    cwr = true;
  }
  if (cwr) {
    // Check still in CWR round
    if (snd_una < next_seq) {
      // Multiplicative Decrease
      cwnd_carry = div(
        cwnd_carry.rem+av_up, acks*G*2);
      cwnd -= cwnd_carry.quot;
    } else {
      cwr = false;
    }
  }
}
\end{verbatim}

As in DCTCP, the EWMA continues to be calculated whether or not there is
congestion feedback, but it is only used in the round after there
is actual congestion marking. However, unlike DCTCP, it continues to be applied
per-ACK during the round of CWR rather than just once at the start of CWR. 

CWR state then takes on a meaning that is nearly the opposite of its classical
meaning. It no longer means `congestion window reduced; no further reduction for
a round'. Instead it means `congestion window \emph{reduction} in progress
during this round'. 

Although the motivation for this algorithm was not to prevent the stall caused
by sudden decrease in \texttt{cwnd}, it would probably serve to address this
problem as well. Therefore it should supplant proportional rate reduction
(PRR~\cite{IETF_RFC6937:PRR}), at least when responding to ECN.

Details of \texttt{cwnd} processing are omitted from the pseudocode if they are
peripheral to the proposed changes. For instance:
\begin{itemize}[nosep]
	\item in real code \texttt{cwnd}
	would be prevented from falling below a minimum (default 2 segments);
	\item the slow-start threshold would track reductions in \texttt{cwnd} (but see
	\S\,\ref{prresp_future} for alternative ideas);
	\item as mentioned earlier,
	counting bytes instead of packets could be used to handle ECN markings on
    packets of different sizes;
	\item the value of \texttt{G} is chosen as a power of 2, so that
	multiplication by \texttt{G} can be implemented with a bit-shift.
\end{itemize}

The pseudocode does, nonetheless, inherently attend to details such as loss of
precision due to integer truncation. And note that \texttt{cwnd} is not updated
on the ACK that ends CWR state, because it is updated on the marked ACK that
starts CWR state.

\newpage
\subsection{The Whole AIMD Algorithm}\label{prresp_AIMD}

\begin{verbatim}
On_each_ACK'd_packet {
  acks_ = update_acks();
  // Update EWMA
  av_carry = 
    div(av_carry.rem+av_up, acks_*G);
  av_up += ce_fb - av_carry.quot;

  if (!ce_fb) {
    // Additive Increase
    // AI & MD denoms must be =
    cwnd_carry = divu(
      cwnd_carry.rem[0]+G*2, 
      acks*G*2, 0);
    cwnd += cwnd_carry.quot;
  } else if (!cwr) {
    // Record start of CWR state
    next_seq = snd_next;
    cwr = true;
  }

  if (cwr) {
    // Check still in CWR round
    if (snd_una < next_seq) {
      // Multiplicative Decrease
      cwnd_carry = divu(
        cwnd_carry.rem[1]+av_up, 
        acks*G*2, 1);
      cwnd -= cwnd_carry.quot;
    } else {
      cwr = false;
    }
  }
}
\end{verbatim}
\paragraph{Additive Increase:} For completeness, the pseudocode above 
includes a Reno-like additive increase.
This is intended for periods when the congestion control is close to its
operating point (if it is not, see \S\,\ref{prresp_future}).

Unlike DCTCP (but like Prague~\cite{Briscoe21b:PragueCC-ID}), the proposed
AI algorithm is not suppressed during
CWR state, with the following reasoning. DCTCP-like congestion controls are
designed to induce roughly 2 ECN marks per round trip in steady state, so RTTs
without marks are meant to be rare. Confining increases to periods that are
not meant to happen creates an internal conflict within DCTCP's own design.
Then, the algorithm's only escape is to store up enough decrease rounds to make
space for a compensating period of increase. This has been found to cause
unnecessary queue variation~\cite{Briscoe20a:PragueCC-present}.

Instead, we continue additive increase regardless of CWR state. In place of
suspending additive increase for a whole round, a fractional increase%
\footnote{Calculated per-ACK as in most TCP implementations}, is skipped if an ACK
carries congestion feedback. This thins down the additive increase as congestion
rises (as recommended in \S\,3.1 of \cite{Briscoe17a:CC_Tensions_TR}).

Unlike DCTCP and unlike Prague, the increase in \texttt{cwnd} per round trip is divided
over the actual number of ACKs, not over the congestion window
(the latter would otherwise increase \texttt{cwnd} by 1 segment whether or not
\texttt{cwnd} was fully used; see \S\,\ref{prresp_Advantage_App-Limited}).

\paragraph{Shared Remainder} The additive increase stores its remainder in the
same \texttt{*cwnd\_carry}
variable as the multiplicative decrease. So both numerator and denominator are
scaled up by \texttt{G*2} so that AI uses the same denominator parameter as MD,
otherwise the
upscaling of the carry variable would be different.

For both the AI and the MD, it will be noticed that a different division
function, divu(), has been used to update \texttt{cwnd\_carry}. This allows the
remainder to be used in both directions without relying on signed integers.

Before the division for AI, the numerator is added to the previous remainder.
But, for MD, the numerator needs to be subtracted from the previous remainder.
Using signed integers would halve the available number space. Instead, the
remainder is always kept positive in the range \texttt{[0, denom-1]}. This is
achieved by maintaining two positive remainders in a two-element array,
\texttt{rem[2]}. Then \texttt{rem[0]} is used as the remainder for an increase
in \texttt{cwnd}, while \texttt{rem[1]} is used for a decrease.

The unsigned division function, \texttt{divu()}, is defined below. It wraps the
kernel macro \texttt{do\_div()}, which is used for efficient integer division.
\texttt{do\_div()} already returns the positive remainder, and writes the result
of the division into the numerator that it was called with. \texttt{divu()}
returns a structure containing both remainders, also defined below. And it takes
an extra boolean flip parameter to say which of the remainders \texttt{do\_div}
should write into.
\begin{verbatim}
typedef struct {
  int quot;
  int rem[2];
} divu_t;

struct divu_t divu(num, denom, flip)
{
  struct divu_t qr;
  qr.rem[flip] = do_div(num, denom);
  qr.rem[!flip] = denom - qr.rem[flip];
  qr.quot = num;
  return qr;
}
\end{verbatim}

Whenever either remainder changes the other is kept in sync as its complement by
the following relation: \texttt{rem[1] = denom - rem[0]}. For example, after a
division by 1000, if the remainder is 400 for increases, the remainder for
decreases will be 600. This can be thought of as a jump from the bottom to the top
of the denominator number space and a flip to viewing everything in the negative
direction as if it was positive.%
\footnote{The two elements of \texttt{rem} might need to be set in one atomic
	operation, depending on how ACK interrupt handling is implemented.}

This ensures that the final result of a division (\texttt{quot}) is always
rounded down (in the negative direction) before altering \texttt{cwnd}. Within
any division that will be used for a decrease, the \texttt{denom} term added
within \texttt{rem[1]} effectively rounds up, but only to keep everything
positive within the division function---the result used outside the division
function is still rounded down.

%

\subsection{Maintaining \texttt{acks\_}}\label{prresp_acks_}

No known TCP implementation maintains the number of ACKs per round. One method
would be to maintain the following variables in the transport connection's state:
\begin{description}[nosep]
	\item[C.cumacks] Per-connection cumulative counter of the number of ACKs
	received during the connection;
	\item[P.cumacks] Per-packet record of the value of C.cumacks when packet P was
	sent (or retransmitted). An ACK often releases multiple packets, which would
	then all share the same value.
\end{description}
Then, on arrival of each ACK, the following pseudocode would maintain \texttt{acks\_}:
\begin{verbatim}
update_acks() {     // Alternative #1
  C.cumacks++;
  if (new data ACKed or SACKed) {
    for latest SACKed or ACKed pkt P
  acks_ = C.cumacks - P.cumacks;
    for each pkt P released by the ACK
  P.cumacks = C.cumacks;
  }
  return acks_;    
}
\end{verbatim}

Alternatively, other methods could be developed that do not need per-packet
state. For instance, an EWMA \texttt{cov\_avg} of the bytes covered by each ACK
could be maintained,%
\footnote{Probably initialized to SMSS at the start of a connection.}
then divided into \texttt{flight} as follows:
\begin{verbatim}
update_acks() {     // Alternative #2
  // EWMA of bytes covered by each ACK
  cov_avg += 
    (acked_or_sacked_bytes - cov_avg)/G2;
  acks_ = flight / cov_avg;
  return acks_;
}
\end{verbatim}
If maintained in the kernel, integer division would be used, and \texttt{flight}
would need to be increased by \texttt{cov\_avg/2} to avoid rounding bias.
Otherwise, floating point arithmetic could be used.

Alternative \#1 calculates the precise value of \texttt{acks\_} over the
previous round. If the gain of the EWMA in alternative \#2 were set so that,
say, \texttt{G2 = 32}, the occasional ACK covering an anomalous amount of data
would have little impact, but \texttt{acks\_} would still immediately pick up
variations in \texttt{flight}.

\paragraph{Receiver-side maintenance of \texttt{acks\_}}
For a peer that is predominantly receiving,not sending,%
\footnote{In the Linux DCTCP implementation, when sending stops, the per-RTT
	clock of the EWMA also stops.} alternative \#2 would not be suitable. For a
receiver, each arriving data packet is also an ACK, so it would clock the EWMA
of marks and (if it fed back congestion on pure
ACKs~\cite{Bagnulo16a:ecn-tcp-ctrl_ID}) it would drive the multiplicative
decrease of \texttt{cwnd}. But it would acknowledge no new data. So
\texttt{flight / cov\_avg} would become \texttt{0/0}.

In the pseudocode for alternative \#1, a predominant receiver would skip the
`if' block and just return the value of \texttt{acks\_} unchanged. This would
probably be sufficient to maintain its EWMA and \texttt{cwnd}, for if it ever
sent some data. It would either use a default initial value of \texttt{acks\_}
or, an earlier value if it had calculated \texttt{acks\_} when it sent some data
in the past (some request data in a single packet would have been sufficient).

If desired, it would be possible to add an `else' to the `ìf' block in
alternative \#1 so that a predominant receiver could fall-back to maintaining
\texttt{acks\_} by using timestamps to estimate the
RTT~\cite{Ming-Chit00:TCP_Assym}, and recording the time it sent each ACK (or
one ACK per RTT), so it could look-up \texttt{P.cumacks} of the ACK that
released an incoming data packet.

\section{(Non-)Concerns}\label{prresp_Non-Concerns}

\subsection{Variable Upscaling}\label{prresp_Variable_Upscaling}

In previous upscaled EWMAs, the upscaling factor was constant. Whereas here, the
upscaling by \texttt{acks\_} varies. So, the meaning of the value of the EWMA
seems to vary. For instance, if there are 50 ACKs per round trip, the EWMA will
be upscaled by 50 times.%
\footnote{Always upscaled by \texttt{G} as well, but that is set to one side here.}
But, if \texttt{cwnd} is then halved (say) due to heavy congestion, in the next
round the number of
ACKs will halve to 25. Then, over the next round, the moving average
fed through from one update to the next will gradually become upscaled by 25 --- half
as much as it was previously. This would seem to distort the meaning of the moving
average. 

However, whether this is a problem depends on what the moving average is meant to mean:
\begin{itemize}
	\item If it's meant to mean the upscaled moving average of the proportion of
	marks (\texttt{alpha}) then, once it is downscaled by the current value of
	\texttt{acks\_}, its meaning will not track \texttt{alpha} exactly unless the
	window remains steady;
	
	\item If it's meant to mean the moving average of the number of marks per round
	(\texttt{v}), then changes in the window will only have second-order effect on
	its meaning because, as the number of updates per round reduces, it is correct
	to upscale each update by less.
\end{itemize} 

It is not obvious which is correct. In DCTCP currently, the multiplicative
decrease can be written as: 
\vspace{-12pt}
\begin{verbatim}
    cwnd -= cwnd * ewma(v/cwnd) / 2,
\end{verbatim}
\vspace{-12pt}
whereas the proposed approach is to use
\vspace{-12pt}
\begin{verbatim}
    cwnd -= ewma(v) / 2.
\end{verbatim}
\vspace{-12pt}
So, it could be argued that, in current DCTCP, dividing \texttt{cwnd} by an EWMA
of \texttt{cwnd} distorts the meaning of \texttt{cwnd}.

There does not seem to be a good theoretical answer. So the question is really
whether the new proposal works well in practice.

\subsection{Circular Dependency?}\label{prresp_No_Circular_Dependency}

There seems to be a circular dependency, because \texttt{av\_up} is both
upscaled by \texttt{acks\_} then used to update \texttt{cwnd}, which
determines \texttt{acks}.


Nonetheless, this is only the same as the circular dependency where cwnd is used
to divide the additive increase over the window, so how fast \texttt{cwnd}
increases depends on itself.
In other words, by definition, incremental update algorithms are circular.

See also \S\,\ref{prresp_Variable_Upscaling} for how \texttt{alpha} in DCTCP depends on \texttt{cwnd}.

\subsection{Anomalous ACK Order}\label{prresp_anomalous_acks}

When packets are acknowledged out of order or retransmissions are acknowledged,
both the alternative algorithms for maintaining \texttt{acks\_} in
\S\,\ref{prresp_acks_} are designed to just work, with no exceptions needed. For
instance, if \texttt{acks\_} has been running at 25, but then one packet goes
missing for, say, 4 ACKs, algorithm \#1 will output 24 on each of the four ACKs.
Then once the late packet appears, \texttt{acks\_} will jump to, say, 29 for one
ACK before returning to 25. This sequence will feed into the decrementing side
of the EWMA, while the congestion feedback on the ACKs feeds into the
incrementing side. So, the effect of the congestion feedback on the delayed
packet will hardly be impacted by the reordering.

\subsection{Biased ACK Coverage}\label{prresp_biased_acks}

When the transition from per-RTT to per-ACK EWMA was explained in
\autoref{tab:invariant}, the number of marks per-ACK (third row) was used, in
preference to the proportion of marks per-ACK (second row). The fraction of
marks per-ACK means, for example, that 1 mark on an ACK covering 4 packets would
be fed into the EWMA as \(\sfrac{1}{4}\). However, to derive the average marking
fraction requires the average of the numerators to be divided by the average of
the denominators, which is not the same as the average of the fractions (unless
all the denominators are the same). Put another way, using the fraction of marks
per-ACK would give more weight to the feedback in ACKs covering fewer packets.
This would also result in bias if the receiver's policy was to trigger an ACK on
receipt of a CE mark.

In contrast, using the number of marks per-ACK does not give more weight to ACKs
that cover fewer packets, because the maximum number of marks on such ACKs is
proportionately smaller. There is a slight inverse correlation between the
coverage of an ACK and the number of ACKs per round. But this only slightly
alters the decay rate. So any bias is doubly slight.

\subsection{Stale Remainder Scaling?}\label{prresp_Stale_Remainder_Scale}

The denominator of each division used to calculate the remainder is not constant
because it contains \texttt{acks\_}. Therefore, if \texttt{acks\_} is growing, a
remainder calculated using earlier values of \texttt{acks\_} will be smaller
than it ought to be. And vice versa if \texttt{acks\_} is shrinking. 

This is a second order effect that should not alter the steady state that the
AIMD converges to; it only puts a slightly different curve on the path to reach
the steady state. 

However, it is important to keep the two remainders in sync straight after
either one has been updated. Otherwise if the complementary remainder is
calculated later, the denominator could have changed, leading to possible subtle
underflow bugs (or more complex code to catch these cases).

\subsection{Advantage to App-Limited Flows?}\label{prresp_Advantage_App-Limited}

The repetitive reduction of \texttt{cwnd} is divided over \texttt{acks} updates in a
round of CWR, so each reduction is scaled down by \texttt{acks} in the
denominator of the call to \texttt{divu()}. This means that \texttt{cwnd} is actually 
decreased by a multiplicative factor of the actual packets in 
\texttt{flight}, not of \texttt{cwnd}.

If a flow is not application-limited, the two amount to the same thing. But for
app-limited flows, \texttt{flight} can be lower than \texttt{cwnd}, so the
reduction in \texttt{cwnd} will be lower.

If a flow is only using a fraction of its congestion window, but it is still
experiencing congestion, there is an implication that other flow(s) must have
filled the capacity that the app-limited flow is `entitled' to but not using.
Then, it could be argued that the other flows have a higher \texttt{cwnd} than
they are `entitled' to, so that the app-limited flow can reduce its
`entitlement' (\texttt{cwnd}) less than these other flows in response to
congestion.

If this argument is not convincing, the reduction in \texttt{cwnd} could be
scaled up by \texttt{cwnd/flight}. However, it is believed that the code is
reasonable, perhaps better, as it stands.

Similar arguments can be used to motivate additive increase over the actual
number of packets in flight, rather than the potential number in flight (the
congestion window).
\section{Evaluation Plan}\label{prresp_Evaluation}

For research purposes, we ought not to introduce more than one change at once, without
evaluating each separately. Therefore, initially, we ought to use the
continually updated EWMA, \texttt{av\_up} to reduce \texttt{cwnd} in the
classical way. That is, on the first feedback of a CE mark, reduce \texttt{cwnd}
once by \texttt{av\_up/(2*G)}. Then suppress further response for a round (CWR
state). This should remove one round of lag (originally spent accumulating the
marking fraction), but not the rest (spent reducing \texttt{cwnd} in response to a single
mark, then doing nothing for a round while the extent of marking is becoming
apparent).

There are even two changes at once in the proposal to update the EWMA per ACK
and to use variable upscaling. It might be possible to evaluate each of these
two innovations separately as well, although they are harder to separate.

On the same principle of one change at a time, A-B testing ought to use Prague
not DCTCP as the base for comparisons, given Prague includes fixes to known
problems with DCTCP's EWMA.%
\footnote{'Bugs' in the implementation of the EWMA in DCTCP have been fixed in
	TCP Prague. Prior to a 2015
	patch~\cite{shewmaker15:Linux_DCTCP_EWMA}, the integer arithmetic for the EWMA
	in Linux floored at 15/1024, which ensured a minimum window reduction even after
	an extended period without congestion marking. That patch toggled the EWMA to
	zero whenever it tried to reduce \texttt{alpha} below that floor, and remained
	unresponsive until congestion was sufficient to toggle \texttt{alpha} back up to
	16/1024. The TCP Prague reference implementation maintains the EWMA in an
	upscaled variant of \texttt{alpha}, as well as using higher precision and
	removing rounding bias.} %
But it might also be interesting to compare this with the pre-2015
implementation of DCTCP that might have accidentally been more responsive in
many scenarios. This would be possible without winding all the code back to that
date by simply setting a floor for \texttt{alpha} at 15/1024 in the Prague code.

The actual comparison should focus on how quickly a Prague flow in congestion
avoidance can reduce in response to a newly arriving flow or a reduction in
capacity.

Pacing should be enabled in both A and B tests. But, initially, segmentation
offload should be disabled in both, to simplify interpretation of the results before
enabling offload in both A and B tests together.

Initially the same gain as DCTCP and Prague (1/16) ought to be used. But it is possible
that the reason DCTCP's gain had to be so low was because of the 1--2 rounds of
built in lag in the algorithm. Therefore, it will be interesting to see if the
gain can be increased (from 1/16 to 1/8 or perhaps even higher), although this
will impact Prague's ability to leave headroom for short flows based on the
recent traffic pattern.%
\footnote{Also see the calculation in
\autoref{prresp_approx} of the difference between per-round and per-ACK EWMAs as
gain increases.}

The thinking here is that a fixed amount of lag in a response is not the same as
smoothing (see \autoref{fig:lag-schematic}). Lag applies the same response but
later. Smoothing spreads the response out, adding lag to the end of the
response, but not to the start. Given
every DCTCP flow's response to each change has been lagged by 1--2 rounds, it is
possible that all flows have had to be smoothed more than necessary, in order to
prevent the excessively lagged responses from causing over-reactions and
oscillations.

One motivation for improving DCTCP's responsiveness is to ensure established
flows yield quickly when new flows are trying to enter the system, without
having to build a queue. However, removing up to two rounds of unnecessary lag
should also help to address the incast problem, at least for transfers that last
longer than one round. Therefore, incast experiments could also be of interest.

It will also be necessary to check performance in the following cases that might
expose poor approximations in the algorithm (relative to Prague):
\begin{enumerate}
	\item When the packets in flight has been growing for some time;
	\item \ldots{}or shrinking for some time;
	\item When the flow is application limited, with packets in flight varying
	wildly, rather than tracking the smoother evolution of \texttt{cwnd}.
\end{enumerate}

\bob{Outcome of the evaluations of version 03 of the algorithm to be added here.}
\section{Related Work}\label{prresp_related}
In 2005 Kuzmanovic~\cite[\S5]{Kuzmanovic05:ECN_SYN_ACK}
presaged the main elements of DCTCP showing that ECN should enable a na\"{\i}ve
unsmoothed threshold marking scheme to outperform sophisticated AQMs like the
proportional integral (PI) controller. It assumed smoothing at the sender, as
earlier proposed by Floyd~\cite{Floyd94:ECN}.

Reducing \texttt{cwnd} in one RTT by half of the marks per round trip
(\texttt{av\_up/2/G}) is similar but not the same as Relentless
TCP~\cite{Mathis09:Relentless}, which reduces \texttt{cwnd} by half a segment on
feedback of each CE-marked packet. The difference is that \texttt{av\_up} is a
moving average, so it does not depend on the number of marks in any specific
round, whereas the Relentless approach does. Relentless was designed for the
classical approach with smoothing in the network, so it immediately applies a
full congestion response without smoothing. In contrast, using the moving
average implements the smoothing in the sender.

Like DCTCP, the per-ACK congestion response proposed in section 5.2 of
\cite{Alizadeh11:DCTCP_Analysis} maintains an EWMA of congestion marking
probability, \texttt{alpha}. But, unlike DCTCP, it reduces \texttt{cwnd} by half
of \texttt{alpha} (in units of packets) on feedback of each ECN mark. This is
partway between Relentless and DCTCP, because it uses the smoothed average of
marking, but it applies it more often in rounds with more marks. This still
causes considerable jumpiness because, with common traffic patterns, marks 
tend to be bunched into one round
then clear for a few rounds. In contrast, the
approach proposed in the present report limits the reduction within any one round
to the averaged number of marks per round (as DCTCP itself does). 

\section{Ideas for Future Work}\label{prresp_future}
\balance
An EWMA of a queue-dependent signal is analogous to an integral controller. It
filters out rapid variations in the queue that do not persist, but it also
delays any response to variations that do persist. Faster control of dynamics
should be possible by adding a proportional element, to create a
proportional-integral (PI) controller within the sender's congestion control.
The proportional element would augment any reduction to the congestion window
dependent on the rate of increase in \texttt{av\_up}.

Separately, it would be possible to use the per-ACK EWMA of marks per round
(\texttt{av\_up/G}) as a good indicator of whether a flow has lost its closed-loop
control signal, for instance because another flow has left the bottleneck, or
capacity has suddenly increased. A flow could then switch into a mode where it
searches more widely for a new operating point, for instance using paced
chirping~\cite[\S\,3]{Misund19a:Paced_Chirping_Linux}. To deem that the closed
loop signal had significantly slowed, it might calculate the average distance
between marks implied by the EWMA \texttt{av\_up}, multiply this by a heuristic
factor, then compare this with the number of packets since the last mark.
Alternatively, it might detect when the EWMA of the marks per round had reduced
below some absolute threshold (by definition, the marks per round of a scalable
congestion control in steady state should be invariant for any flow rate).

\newpage
\addcontentsline{toc}{section}{References}

{%
\scriptsize%
\bibliography{responsiveness}}

\clearpage
\appendix
\section{Approximations}\label{prresp_approx}

The per-ACK EWMA is not intended to mimic a per-RTT EWMA. Otherwise, the per-ACK
EWMA would have to reach the same value by the end of the round, irrespective of
whether markings arrived early or late in the round.
It is more important for the EWMA to quickly accumulate any markings early in
the round than it is to ensure that the EWMA reaches precisely the same value by
the end of the round. 

Neither is it important that a per-ACK EWMA decays at precisely the same rate as
a per-round EWMA (assuming they both use the same gain). The gain is not
precisely chosen, so if a per-ACK EWMA decays somewhat more slowly, it is
unlikely to be critical to performance (if so, a higher gain value can be
configured).

However, it \emph{is} important that a per-ACK EWMA decays at about the same
rate however many ACKs there are per round, although the decay rate does not
have to be precisely the same.

The per-ACK approach uses the approximation that one reduction with gain \(1/G\)
is roughly equivalent to \(n\) repeated reductions with \(1/n\) of the gain.
Specifically, that \((1 - 1/nG)^n \approx 1 - 1/G\).

\begin{align*}
(1 - 1/nG)^n &=       1 + \frac{n}{-nG} + \frac{n(n-1)}{2(-nG)^2} + \ldots \\
             &=       1 - \frac{1}{G} + O\left(\frac{1}{G^2}\right)\\
             &\approx 1 - \frac{1}{G}
\end{align*}

\begin{figure}[h]
	\includegraphics[width=\linewidth]{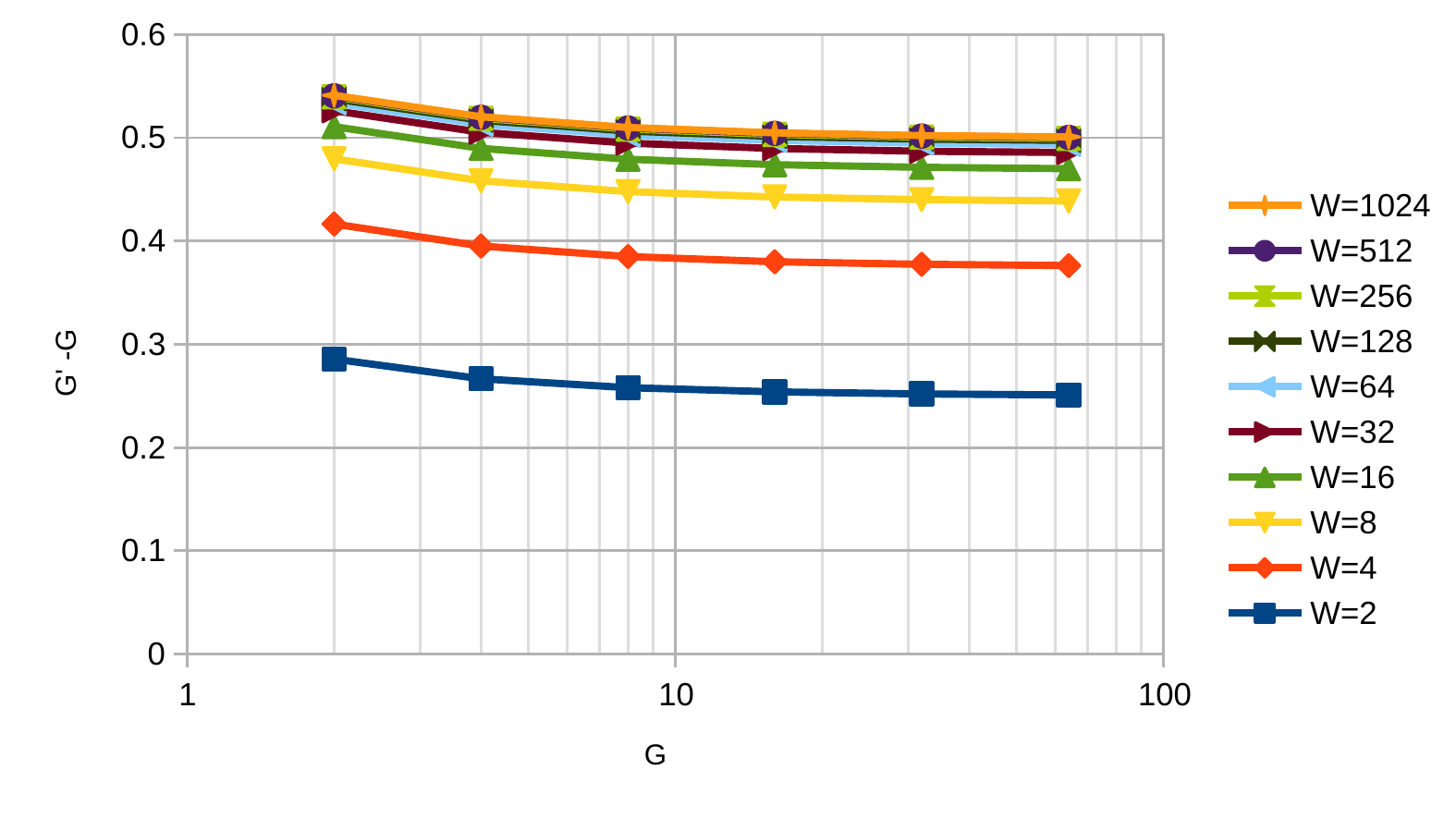}
	\caption{Difference between gain used for multiple per-ACK reductions, \(G^\prime\), and the gain of one equivalent reduction, \(G\).}
	\label{fig:per-ack-approx}
\end{figure}

To quantify the error, we define the effective gain (\(1/G^\prime\)) as the
per-RTT gain that would give an equivalent reduction to multiple smaller per-ACK
reductions using the original gain (\(1/G\)). Numerically, we find that
\(G^\prime \approx G + 1/2\) (see \autoref{fig:per-ack-approx}).

For instance, multiple reductions with \(G\approx15.5\) are roughly equivalent to one
reduction with \(G^\prime=16\).

This can be explained because most of the error comes from omission of the
\(O(1/G^2)\) term. So we set
\begin{align*}
1-\frac{1}{G^\prime} &\approx 1 - \frac{1}{G} + \frac{(n-1)}{2nG^2},
\intertext{which, in the worst case of large \(n\), reduces to}
G^\prime &\approx \frac{2G^2}{(2G-1)}.
\intertext{Then the difference between the reciprocals of the effective and
	actual gains is}
G^\prime - G &\approx \frac{G}{(2G-1)}
\end{align*}
Other than for low values of \(G\) this difference is indeed roughly \(1/2\). 

The worst-case error occurs when \(G\) is small and \(W\) is large. Multiple
reductions using any high value of \(W\) and the lowest practical value of \(G
(=2)\) would be equivalent to a single reduction using \(G^\prime\approx2.54\)
(i.e.\ the error in this worst-case is about 0.54).

\onecolumn%
\addcontentsline{toc}{part}{Document history}
\section*{Document history}

\begin{tabular}{|c|c|c|p{3.5in}|}
 \hline
Version &Date &Author &Details of change \\
 \hline\hline
00A          &07 Nov 2020&Bob Briscoe &First draft.\\\hline%
00B          &29 Nov 2020&Bob Briscoe &Added \texttt{cwnd} reduction and increase. Defined reusable function \texttt{repetitive\_div()}. Corrected use of \texttt{cwnd} to \texttt{flight}, and distinguished current \texttt{flight}, from \texttt{flight\_} when marks were averaged. Added abstract; schematic of problem; sections on evaluation plan, related work and future work; and appendix on approximations.\\\hline%
00C &02 Dec 2020  &Bob Briscoe &Altered algorithms from hand-crafted repetitive\_div() to div() in stdlib\\\hline%
01           &19 Jan 2021&Bob Briscoe &Added CC to title, altered abstract, added motivation to intro and issued.\\\hline
02 &20 Jan 2021&Bob Briscoe &Improved abstract, intro \& eval'n plan.\\\hline%
03A&07 Feb 2021&Bob Briscoe &More care distinguishing DCTCP/Prague variants. Defined \texttt{div2()} to fix root cause of remainder underflow. Explained shared remainders properly. Changed \texttt{g} terminology to \texttt{G}. Added todo notes incl.\ EMWA init.\\\hline%
03B&06 Aug 2022&Bob Briscoe &Added context and extra schematic in introduction. Rewrote intuition section. New algorithm upscaling by \texttt{acks\_} not \texttt{flight\_} and feeding in the amount of CE marked packets (or bytes), not the proportion; to allow for ACKs covering multiple packets.\\\hline%
03 &7 Aug 2022 &Bob Briscoe &Minor updates and corrections throughout.\\\hline%
\metaversion &\metadate    &Bob Briscoe &Added \S\,\ref{prresp_acks_} on maintaining \texttt{acks\_}. Changed title from ``Improving DCTCP/Prague Congestion Control Responsiveness''. Clarified and edited throughout, including rearrangement of first two sections.\\\hline%
\hline%
\end{tabular}

\end{document}